\documentclass[a4paper]{article}

\usepackage{INTERSPEECH2018}

\usepackage{CJKutf8}
\usepackage{color}
\usepackage{amsmath,graphicx}
\usepackage{amsfonts} 
\usepackage{multirow}

\usepackage{hyperref}

\usepackage[table]{xcolor}

\def\first{{1\raise.5ex\hbox{\small st }}}
\def\second{{2\raise.5ex\hbox{\small nd }}}
\def\rd{{\raise.5ex\hbox{\small rd }}}
\def\th{{\raise.5ex\hbox{\small th }}}

\def\TER{{11.7}}

\title{Tone recognition using lifters and CTC}
\name{Loren Lugosch, Vikrant Singh Tomar}
\address{
  Fluent.ai Inc., Montr\'eal, Qu\'ebec, Canada}
\email{loren.lugosch@fluent.ai, vikrant@fluent.ai}

\begin{document}

\maketitle
\begin{abstract}
  In this paper, we present a new method for recognizing tones in continuous speech for tonal languages. The method works by converting the speech signal to a cepstrogram, extracting a sequence of cepstral features using a convolutional neural network, and predicting the underlying sequence of tones using a connectionist temporal classification (CTC) network. The performance of the proposed method is evaluated on a freely available Mandarin Chinese speech corpus, AISHELL-1, and is shown to outperform the existing techniques in the literature in terms of tone error rate (TER).

\end{abstract}

\section{Introduction}
\label{sec:intro}
Tones are an essential component of the phonology of many languages. A tone is a pitch pattern that distinguishes or inflects words. For example, in Mandarin Chinese, the words for ``mom'' (\begin{CJK*}{UTF8}{gbsn}妈\end{CJK*} mā),
``hemp'' (\begin{CJK*}{UTF8}{gbsn}麻\end{CJK*} má),
``horse'' (\begin{CJK*}{UTF8}{gbsn}马\end{CJK*} mǎ), and
``scold'' (\begin{CJK*}{UTF8}{gbsn}骂\end{CJK*} mà) 
are composed of the same two phones and are distinguishable only through their tones. Consequently, automatic speech recognition systems for tonal languages cannot rely on phones alone and must incorporate some sort of tone recognition to avoid ambiguity.

Modern speech recognition systems commonly use spectral features, such as Mel-filtered cepstrum coefficients (MFCCs), Mel-filterbank features (FBANKs), and perceptual linear prediction features (PLP), as input features. These representations work well for phone recognition, but do not carry information about pitch. Therefore, state-of-the-art speech recognizers for tonal languages append a set of pitch features to each spectral feature vector in order to recognize tones \cite{Liu2015}. These pitch features are typically an estimate of the fundamental frequency ($F_0$) and the probability-of-voicing (PoV) for each frame \cite{Ghahremani2014}. In this work, we refer to this class of features as ``hard-decision pitch features'' (HDPFs).

While the use of HDPFs has been found to improve both tone error rate (TER) and word error rate (WER) or character error rate (CER) for tonal language recognition \cite{metze2013models}, we believe that models that use HDPFs may ignore other important aspects of tonal speech useful for discrimination. This can be argued by analogy from phone recognition. It is known from acoustic phonetics that formant frequencies are the dominant features used to distinguish vowels and certain consonants \cite{ladefoged2014course}. However, phone recognizers typically do not explicitly estimate formant frequencies; rather, they fit a model to the spectral features, thereby implicitly learning about formants but also about other aspects of the input relevant for discriminating phones. Likewise, although the fundamental frequency is the dominant auditory feature used by humans for discrimination of tones, a model that exploits all the available information in the input could provide better tone recognition accuracy.

In this paper, we present a new method for recognizing tones. Unlike the aforementioned HDPF-based approaches, our method does not explicitly estimate the pitch of a speech signal; rather, it uses a trainable front-end based on a convolutional neural network (CNN) that takes as input a cepstrogram and outputs cepstral features. We refer to these features as ``lifter features'' (LFs) because they are obtained by ``liftering'' the input in the ``quefrency'' domain. The sequence of LFs is fed into a recurrent neural network (RNN) with connectionist temporal classification (CTC) to predict a sequence of tones. The trainable feature extractor and CTC network together form a single neural network that can be trained using stochastic gradient descent. 

The proposed tone recognizer offers a number of advantages over existing approaches: it is conceptually simple, can be trained in an end-to-end fashion without obtaining frame-level tone labels through forced alignment or manual labelling, and uses features that are optimized for discrimination rather than features derived from linguistic intuition. The experimental results presented in Section \ref{sec:experiments} show that the proposed system attains a tone error rate of \TER \%. To our knowledge, this is the best error rate reported for the task of tone recognition in continuous Mandarin speech.

Another important aspect of this work is our use of a freely available large vocabulary tonal language corpus. The existing papers on the topic of tone recognition often use expensive tonal language datasets, such as HKUST/MTS \cite{Fung2006} and CALLHOME \cite{Canavan1997}, or in-house datasets not available to the public. This is problematic for researchers, since it is difficult to make objective comparisons between different approaches without obtaining these datasets. In the spirit of reproducible research, we perform our experiments using AISHELL-1, a freely available dataset of spoken Mandarin Chinese intended for developing large vocabulary continuous speech recognition (LVCSR) systems \cite{Bu2017}.

The rest of the paper is organized as follows. Section \ref{sec:background} provides background on the problem of tone recognition. Section \ref{sec:proposed_method} describes our model and justifies the design choices listed above. Section \ref{sec:experiments} describes an experiment to validate the proposed method. Finally, Section \ref{sec:conclusion} concludes with some ideas for future work.

\section{Background}
\label{sec:background}
This section formally defines the problem of tone recognition and reviews some of the existing approaches to solving it. 

\subsection{Problem statement}
\label{ssec:problem_statement}
Tone recognition can be treated as a sequence prediction problem. Let $\mathcal{A}$ be an alphabet of tones, $X = \{X_t \in \mathbb{R} \mid t = 1,\dots,T\}$ be a speech signal, and $Y = \{Y_u \in \mathcal{A} \mid u = 1,\dots,U\}$ be the tone sequence underlying the speech signal. A tone recognizer computes $\hat{Y} = f(X)$, an estimate of $Y$ given the input signal.

The metric of interest, tone error rate (TER), is defined as the average Levenshtein distance, $LD$, between $\hat{Y}$ and $Y$, 

\begin{equation}
    LD(\hat{Y}, Y) = \frac{I + D + S}{U},
\end{equation}
where $I$ is the number of insertions, $D$ is the number of deletions, and $S$ is the number of substitutions. The problem we are interested in solving is finding a function $f(X)$ that minimizes TER. 

\begin{figure}[t]
    \centering
    \includegraphics[width=0.8\linewidth]{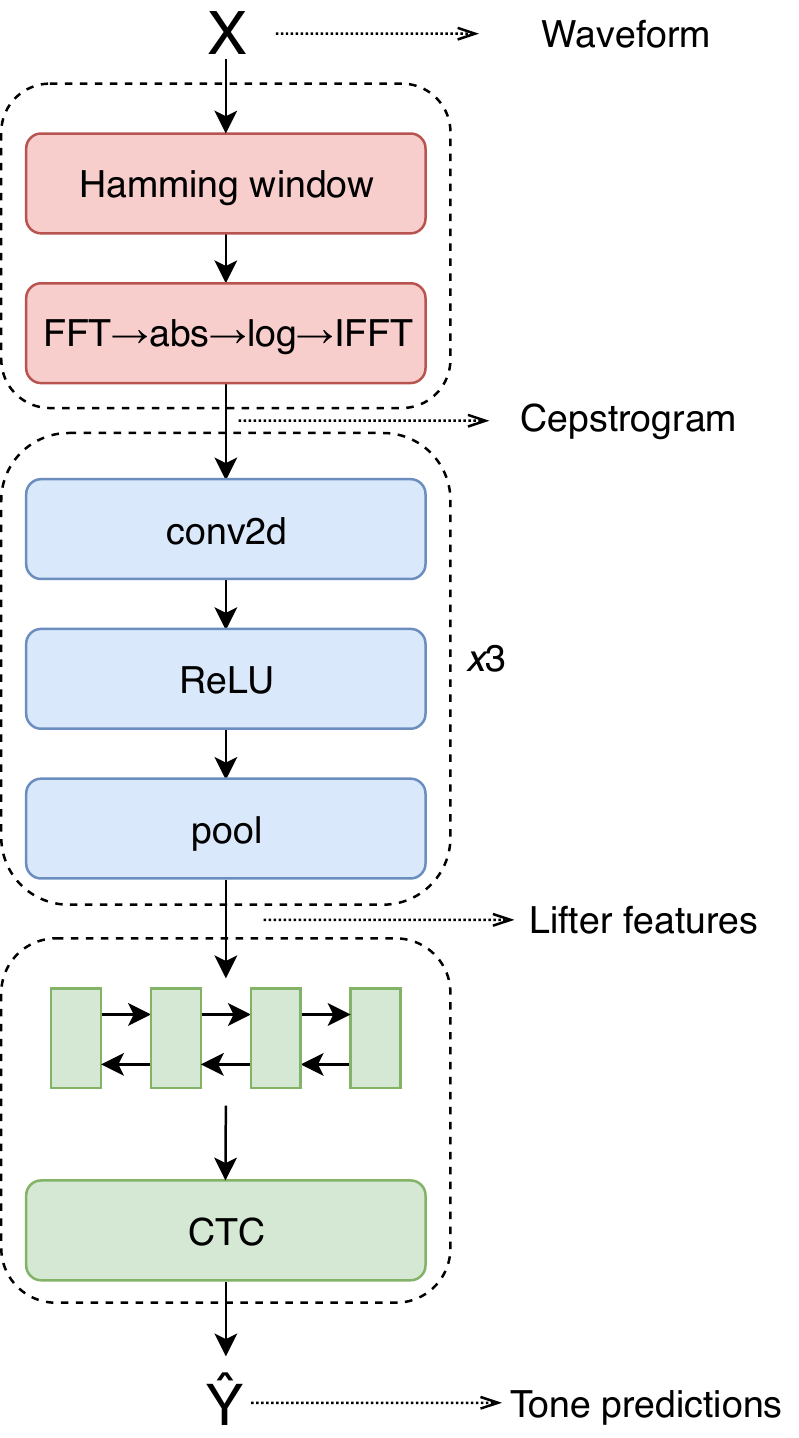}
    \caption{Block diagram of the proposed recognizer.}
    \label{fig:arch}
\end{figure}

\subsection{Existing approaches}
\label{sec:related_work}

Conventional approaches treat hard-decision pitch features (HDPFs) as sufficient statistics for tone recognition. For example, Huang \textit{et al.}\ \cite{Huang2000} use $F_0$ with utterance-level normalization, delta-$F_0$, and degree of voicing as the input to a Gaussian mixture model (GMM)-based recognizer. Lei \textit{et al.}\ \cite{Lei2006} use the $F_0$ contour and syllable duration as the input to a multilayer perceptron (MLP). Some approaches also provide the tone model with MFCCs or FBANKs, such as the RNN-based tone recognizer in \cite{Huang2017} and \cite{adams2017phonemic}, and speech recognizers that model tones as components of tonal phones (see e.g.\ \cite{Liu2015}, \cite{Bu2017}, \cite{metze2013models}).

In \cite{Ryant2014}, Ryant \textit{et al.}\ train two MLP-based tone recognizers, one that used $F_0$ and amplitude alone as input features and one that used MFCCs alone as inputs. Remarkably, they report that the MFCC-based recognizer handily outperforms the HDPF-based recognizer. Similarly, in \cite{chen2016tone}, Chen \textit{et al.}\ train a convolutional network to take as input a window of MFCCs for a single tonal syllable and predict its tone. Although $F_0$ can be estimated very accurately, these results show that $F_0$-based features are not the best features for tone recognition, or at least that there is some information in the input signal that is lost when HDPFs alone are used.

Some papers have suggested extracting alternative features from the input signal. In \cite{li2011improved}, Li \textit{et al.}\ convert the input signal to a spectrogram and convolve it with a set of Gabor filters to yield a set of feature maps. Frame-level tone labels are obtained using forced alignment, and an MLP is trained to predict the tone label for each frame using the feature maps. Likewise, Kalinli \cite{Kalinli2011} uses a biologically inspired system of Gabor filters to extract features from the spectrogram. Baidu's Deep Speech 2 recognizer maps the raw spectrogram directly to Chinese characters, using many convolutional layers and recurrent layers to extract pitch features implicitly \cite{amodei2016deep}.

There are two main issues with conventional tone recognition approaches. First, they use derivative features that may throw away information useful for tone recognition. This issue is somewhat addressed in the latter papers, which use the raw spectrogram as input. However, we believe that the \textit{cepstrogram} \cite{cepstrum} is the appropriate input from which features should be learned, rather than the spectrogram. Second, conventional methods segment the input so that the tone label is known for each frame. This requires either forced alignment, in which a separate model is trained to segment the input, or manual labelling, which is tedious and expensive. A simpler approach is to use an end-to-end method with a sequence-level training criterion. In the next section, we describe a new method for tone recognition that makes use of these two solutions.

\section{Proposed method}
\label{sec:proposed_method}
The proposed tone recognizer is a neural network composed of a preprocessing module, a convolutional network, and a recurrent network (Fig.\ \ref{fig:arch}). In this section, we give more details on each of these components.

\subsection{Preprocessing module}
\label{sec:preprocessing_network}
\begin{figure}[t]
    \centering
    \includegraphics[width=\linewidth]{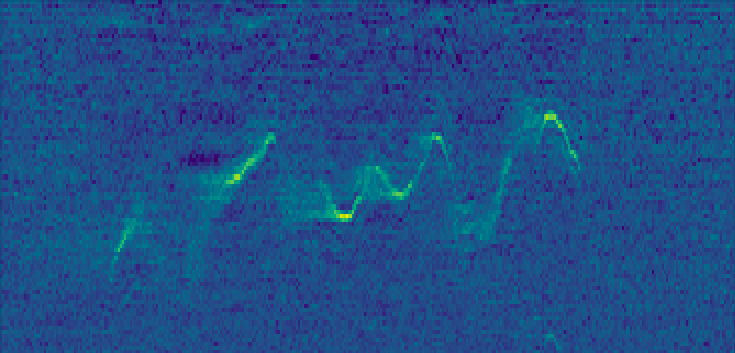}
    \caption{Pitch trace visible in cepstrogram.}
    \label{fig:cepstrogram}
\end{figure}

To a first approximation, tones in speech are pitch trajectories.  Thus, a natural way to approach the problem of tone recognition would be to look for patterns in a 2D time-frequency ``pitch map'' of the input. The cepstrogram is a good candidate for such a 2D representation because i) like the spectrogram, the cepstrogram contains all the information present in the speech signal (except phase), and at the same time ii) the pitch of a voice appears as a single peak at each timestep. Fig.\ \ref{fig:cepstrogram} shows how the pitch of the voice appears clearly in a section of the cepstrogram.

To obtain the cepstrogram, the input signal is split into short, overlapping frames, and each frame is multiplied by a windowing function. Next, the cepstrum is computed for each windowed frame. The cepstrum is defined as the inverse discrete Fourier transform of the spectrum,
\begin{equation}
    \text{cepstrum}(x) = \text{IDFT}(\log |\text{DFT}(x)| ),
\end{equation}
where $x$ is a windowed frame of $X$. We do not apply Mel filterbanks because they smooth the spectrum, which makes the periodic peaks of the spectrum less obvious; instead, we use the raw spectrum to compute the cepstrum. The cepstrogram is then simply the concatenation of all the cepstra.

\subsection{Convolutional network}
\label{sec:convolutional_network}

The identity of a tone is not changed if the overall pitch trajectory of the utterance is translated in time or frequency, in the same way that a melody is not changed if it is sung at a different time or in a different key. Therefore, we use convolutional layers to extract translation-invariant patterns from the input cepstrogram. We use three layers of lifters with ReLU activations and max-pooling. The pooling allows the input to be aggressively downsampled in both time and quefrency, improving the translation-invariance of the feature detection and reducing the length of the sequences the subsequent stage must learn to process. The lifter features (LFs)---the feature maps of the last convolutional layer---are stacked for each timestep to obtain a single 2D feature map, which is presented to the recurrent network.

\subsection{Recurrent network}
\label{sec:training}
We use a sequence-to-sequence model---an RNN with CTC \cite{CTC}---to translate the LFs into tones. This avoids the additional arduous step of aligning and segmenting the input into labelled sections mentioned above. In principle, any sequence-to-sequence model could be used; we chose to use a CTC model because of its simplicity. CTC is ideal for modelling sequences of events in which the same event may occur multiple times consecutively, as is the case for tone recognition. 

\section{Experiments}
\label{sec:experiments}
This section describes the experimental setup and dataset. A number of different recognizers are tested to compare with the proposed method. The experiments are conducted on a large vocabulary Mandarin corpus.

\subsection{Setup}
\label{ssec:setup}
The experiments are performed on the AISHELL-1 dataset.\footnote{The dataset can be downloaded for free at \url{http://www.openslr.org/33/}} AISHELL-1 consists of 165 hours of clean speech recorded by 400 speakers from various parts of China, 47\% of whom were male and 53\% of whom were female. The speech was recorded in a noise-free environment, quantized to 16 bits, and resampled to 16,000 Hz. The training set contains 120,098 utterances from 340 speakers (150 hours of speech), the dev set contains 14,326 utterances from 40 speakers (10 hours), and the test set contains 7,176 utterances from the remaining 20 speakers (5 hours). 

Table \ref{config} lists the hyperparameters used in the LF-based recognizer for these experiments. We used a bidirectional gated recurrent unit (BiGRU) \cite{Cho2014} with 128 hidden units in each direction as the RNN. The RNN has an affine layer with 6 outputs: 5 for the 5 Mandarin tones, and 1 for the CTC ``blank'' label.

\begin{table}[t]
\caption{Layers of the proposed recognizer.}
\label{config}
\begin{center}
\begin{tabular}{| l | r |}
\hline
\bf{Layer type}  &\bf{Hyperparameters}\\
\hline
framing         & 25 ms w/ 10 ms stride\\
windowing         & Hamming window\\
FFT         & length-512\\
abs         & ---\\
log        & ---\\
IFFT         & length-512\\
conv2d         & 11x11, 16 lifters, stride 1\\
pool         & 4x4, max, stride 2\\
activation         & ReLU\\
conv2d         & 11x11, 16 lifters, stride 1\\
pool         & 4x4, max, stride 2\\
activation         & ReLU\\
conv2d         & 11x11, 16 lifters, stride 1\\
pool         & 4x4, max, stride 2\\
activation         &  ReLU\\
dropout         & 50\%\\
recurrent         & BiGRU, 128 hidden units \\
CTC          &  --- \\
\hline
\end{tabular}
\end{center}
\vspace{-2em}
\end{table}

There are two baseline recognizers used in this work. The first (Baseline 1) is an RNN-CTC recognizer that takes sequences of MFCCs and HDPFs as input. We gave this network a second recurrent layer, with dropout between the two recurrent layers, and used 160 hidden units for each recurrent layer. This gives the network more parameters than the LF-based recognizer, so as to give it a competitive advantage over our proposed approach. The second recognizer (Baseline 2) is identical to the proposed recognizer, but with the first 25 coefficients of each cepstrum set to 0, and the remaining 231 coefficients unchanged. The purpose of Baseline 2 is to investigate how important the ``non-pitch'' aspects of the input signal are for tone recognition. In the cepstrum, the ``low-time'' coefficients contain information about the vocal tract, whereas the ``high-time'' (HT) coefficients contain information about the glottal excitation. Thus, zeroing out the low-time coefficients effectively erases the non-pitch information. We also trained a spectrogram-based recognizer to show that the cepstrogram is more suitable as the input. This recognizer is the same as the recognizer described in Table \ref{config} but with the IFFT step removed so that the preprocessing module computes a spectrogram rather than a cepstrogram. However, the spectrogram-based recognizer was unable to learn.

We used the Kaldi \cite{kaldi} recipe for AISHELL-1 to prepare the corpus. We used the Kaldi scripts to compute 13-dimensional MFCCs and 3-dimensional HDPFs to be used in Baseline 1 and normalized each of these features on a per-utterance basis to have zero mean and unit variance. To train the recognizers, we extracted the tone labels and used the standard CTC loss function, $-\log p(Y|X)$, in stochastic gradient descent. The networks were each trained for 20 epochs using the Adam optimizer \cite{ADAM} with an initial learning rate of 0.001 and gradient clipping. The learning rate was halved at the end of an epoch if the loss on the dev set was higher than in the previous epoch. We used the SortaGrad curriculum learning strategy \cite{amodei2016deep} in which training sequences are drawn from the training set in order of length during the first epoch and randomly in subsequent epochs. Decoding the tone sequences using a beam search with a very wide beam yielded TERs that were lower than using simple greedy decoding by only 0.1\% for all recognizers; here, we report only the results of greedy decoding.

\subsection{Results}
\label{ssec:results}
The performance of a number of tone recognizers is compared in Table \ref{comparison}. In the upper half of the table, we list other Mandarin tone recognition results reported elsewhere in the literature.  In the lower half of the table, we list the results of the LF-based recognizer and the two baseline recognizers described in Section \ref{ssec:setup}. Our proposed approach achieves better results than our Baselines 1 and 2 and the other reported results by a wide margin, with a TER of \TER\%. We acknowledge that it is not entirely fair to compare our results with the previously published results, as they are based on different datasets. Our Baseline 1 is most directly comparable to the recognizer in \cite{Huang2017}, since both use MFCCs and HDPFs as the input to an RNN; however, the RNN in \cite{Huang2017} processes the audio of each syllable separately to classify its tone, whereas our RNN processes the audio for the entire utterance and outputs the sequence of tones. Thus, our recognizer has to learn to solve a more difficult problem because the locations of the individual tones are not provided to the system---it is able to learn only from the sequence of tone labels.

The results shown in Table \ref{comparison} support our claim that the cepstrogram is a better choice of input for the tone recognizer. Additionally, there is a significant gap between the performance of the high-time cepstrogram-based recognizer (Baseline 2) and that of the recognizer which uses the entire cepstrogram (proposed). This suggests that ``non-pitch'' features are indeed useful for tone recognition, as some researchers have hypothesized \cite{Ryant2014}. As mentioned earlier, our spectrogram-based recognizer was entirely unable to learn, so we have not included it in the table. We are not claiming that it is impossible to learn good features in the spectrogram; for example, Kalinli's work listed in Table \ref{comparison} uses the spectrogram as the input, so perhaps our spectrogram-based recognizer could be made to learn by tuning the hyperparameter settings. However, one should expect that the spectrogram is more challenging to learn from, since the pitch does not appear as a localized peak, but rather as a global pattern of harmonics. We have not explored tuning the spectrogram-based recognizer.

\begin{table}[t]
\caption{Comparison of recognition results.}
\vspace{-2em}
\label{comparison}
\begin{center}
\begin{tabular}{| l | l | l |}
\hline
\bf{Method} & \bf{Model and input features} & \bf{TER}\\
\hline
Lei \textit{et al.}\ \cite{Lei2006} & HDPF $\rightarrow$ MLP & 23.8\% \\
Kalinli \cite{Kalinli2011} & Spectrogram $\rightarrow$ Gabor $\rightarrow$ MLP & 21.0\% \\
Huang \textit{et al.}\ \cite{Huang2000} & HDPF $\rightarrow$ GMM & 19.0\% \\
Huang \textit{et al.}\ \cite{Huang2017} & MFCC + HDPF $\rightarrow$ RNN & 17.1\% \\ 
Ryant \textit{et al.}\ \cite{Ryant2014} & MFCC $\rightarrow$ MLP & 15.6\% \\
\hline
Baseline 1 & MFCC + HDPF $\rightarrow$ RNN + CTC &  18.1\% \\
Baseline 2 & HT CG $\rightarrow$ CNN $\rightarrow$ RNN + CTC         & 15.1\% \\
Proposed & CG $\rightarrow$ CNN $\rightarrow$ RNN + CTC          & \bf{\TER\%} \\
\hline
\end{tabular}
\end{center}
\vspace{-2em}
\end{table}

\begin{table}[]
\centering
\caption{Breakdown of errors.}
\label{breakdown}
\begin{tabular}{|l|r|r|r|}
\hline
           & \multicolumn{1}{c|}{\textbf{Insertions}} & \multicolumn{1}{c|}{\textbf{Deletions}} & \multicolumn{1}{c|}{\textbf{Substitutions}} \\ \hline
Baseline 1 & \textbf{467}                             & 4934                                    & 31459                                       \\ \hline
Proposed   & 544                                      & \textbf{1382}                           & \textbf{21854}                              \\ \hline
\end{tabular}
\end{table}

\begin{table}[t]
\centering
\caption{Per-tone accuracy.}
\label{tone_accuracy}
\begin{tabular}{|l|c|c|c|c|c|}
\hline
            & \multicolumn{1}{c|}{\textbf{Tone 0}} & \multicolumn{1}{c|}{\textbf{Tone 1}} & \multicolumn{1}{c|}{\textbf{Tone 2}} & \multicolumn{1}{c|}{\textbf{Tone 3}} & \textbf{Tone 4}    \\ \hline
Baseline 1 & \textbf{77.2\%}                   & 81.7\%                            & 88.1\%                            & 69.5\%                            & 85.7\%          \\ \hline
Proposed         & 73.6\%                            & \textbf{90.6\%}                   & \textbf{88.9\%}                   & \textbf{82.9\%}                   & \textbf{91.9\%} \\ \hline
\end{tabular}
\end{table}

Table \ref{breakdown} gives a more detailed analysis of the types of errors the recognizers make. The count for each event was computed using the backtrace of the edit distance between the true tone sequences and the decoded tone sequences. While the proposed recognizer makes slightly more insertion errors, it makes far fewer deletion errors and substitution errors.

Table \ref{tone_accuracy} shows the recognition accuracy for each tone separately. Both recognizers have difficulty detecting tone 0. This is to be expected, since tone 0 in Mandarin is a ``neutral'' tone. Tone 3 is also difficult, although the LF-based recognizer performs significantly better for this tone. We have found that tone 3 is often mistaken for tone 2 due to the phenomenon of tone sandhi. Tone sandhi is a phonological mechanism that makes changes to tones in the context of neighboring tones. For example, a sequence of 3rd tone syllables that would be pronounced $/3,3/$ without tone sandhi becomes $[2,3]$ after the application of tone sandhi in Mandarin. Given a sequence that sounds like $[2,3]$, it is impossible to determine whether the underlying sequence was $/3,3/$ or $/2,3/$, since both hypothetical sequences could have given rise to $[2,3]$. In such cases, it is necessary to use a language model to determine which underlying sequence is more likely.

\subsubsection{Honorable mentions}
\label{ssec:honorable_mentions}

We have listed results in Table \ref{comparison} for recognizers that rely only on acoustic information for continuous speech. Some additional results not listed in the table are worth mentioning. 

First, Lei \textit{et al.}\ in \cite{Lei2006} find that simply extracting the tones from the transcript produced by a full ASR system resulted in a TER of 9.3\%, compared to 23.8\% when using only acoustic information. This is possible because the language model corrects some of the tone errors made by the acoustic model (e.g., ``call my mom'' is more likely than ``call my hemp''). Here, we are interested in studying the performance of the tone model alone. Our tone recognizer is given only tone labels during training, not phone, character, or word labels. 

Second, in \cite{chen2016tone}, Chen \textit{et al.}\ achieve a TER of 4.5\% using a convolutional MFCC-based recognizer. However, their dataset consists of single syllables spoken in isolation. Recognizing tones spoken in isolation is a less challenging task because the speaker produces them more carefully. In fact, it is possible to classify isolated tones fairly accurately simply using the duration of the syllable, whereas tones produced in continuous speech all have roughly the same duration \cite{yang2017duration}. 

\section{Conclusion}
\label{sec:conclusion}
This paper has proposed a new method for tone recognition using the cepstrogram and CTC training.
Using experiments conducted on the AISHELL-1 Mandarin Chinese speech dataset, our method achieves a tone error rate of 11.7\%, which to our knowledge is the best reported tone error rate for continuous Mandarin speech. 
Future work could include incorporating the proposed tone recognizer into a complete tonal language LVCSR system as an alternative to hard-decision pitch features. We believe that this would both improve the character error rate and reduce the number of parameters needed for the acoustic model.

\clearpage
\bibliographystyle{IEEEtran}

\bibliography{main.bbl}

\end{document}